\documentclass[
 reprint,
prl,
]{revtex4-1}

\usepackage{graphicx}
\usepackage{dcolumn}
\usepackage{bm}

\begin{document}

\preprint{PRL}

\title{Direct Single Molecule Imaging of Enhanced Enzyme Diffusion}



\author{Mengqi Xu}
\affiliation{Department of Physics, University Massachusetts Amherst, MA 01003}
\author{Lyanne Valdez}
\affiliation{Department of Chemistry, Pennsylvania State University, State College, PA 18602}
\author{Aysuman Sen}
\affiliation{Department of Chemistry, Pennsylvania State University, State College, PA 18602}
\author{Jennifer L. Ross}
\affiliation{Department of Physics, University Massachusetts Amherst, MA 01003}


\date{\today}

\begin{abstract}
Recent experimental results have shown that active enzymes can diffuse faster when they are in the presence of their substrates. Fluorescence correlation spectroscopy (FCS), which relies on analyzing the fluctuations in fluorescence intensity signal to measure the diffusion coefficient of particles, has typically been employed in most of the prior studies. However, flaws in the FCS method, due to its high sensitivity to the environment, have recently been evaluated, calling the prior diffusion results into question. It behooves us to adopt complementary and direct methods to measure the mobility of enzymes in solution. Herein, we use a novel technique of direct single-molecule imaging to observe the diffusion of single enzymes. This technique is less sensitive to intensity fluctuations and gives the diffusion coefficient directly based on the trajectory of the enzymes. Our measurements recapitulate that enzyme diffusion is enhanced in the presence of its substrate and find that the relative increase in diffusion of a single enzyme is even higher than those previously reported using FCS. We also use this complimentary method to test if the total enzyme concentration affects the relative increase in diffusion and if enzyme oligomerization state changes during catalytic turnover. We find that the diffusion increase is independent of the total background concentration of enzyme and the catalysis of substrate does not change the oligomerization state of enzymes.

\begin{description}
\item[PACS numbers]
82.39.−k, 87.16.Uv, 82.60.Hc
\end{description}
\end{abstract}

\pacs{Valid PACS appear here}
\maketitle



Molecular enzymes are active matter systems that use energy to perform a variety of tasks required for the basic functions of cells. Enzymatic activity is thought to be essential to maintain cellular temperature and active mixing of the crowded and visco-elastic environment inside cells \cite{Guo2014, Parry2014}. When active enzymes were bound to the surface of micron-scale colloidal particles, they were able to self-propel in the presence of the enzyme substrate \cite{Ma2015, Patino2018}. Thus, active enzymes can act as a source of propulsion to move large-scale objects in aqueous media. 

Recent experimental studies have demonstrated that active enzymes diffuse faster in the presence of their corresponding enzymatic substrates \cite{Yu2009, Muddana2010, Sengupta2013, Sengupta2014, Bustamante, Golestanian2017, Jee2017, Gunther2018}. These prior studies measured the relative increase in the diffusion coefficient, ranging from 20\% to 80\%, depending on the enzyme used and the substrate concentration \cite{Yu2009, Muddana2010, Sengupta2013, Sengupta2014, Bustamante, Golestanian2017, Jee2017, Gunther2018}. A major drawback of prior measurements is that they all use a single method: fluorescence correlation spectroscopy (FCS). In FCS, the diffusion coefficient is found by measuring and fitting the autocorrelation function of the fluctuations in fluorescence intensity signal to a diffusion model. Although FCS is referred to as a single-molecule technique, the measurement often relies on signal from several particles \cite{Bacia2014}. Further, the intensity fluctuations of fluorophores are highly sensitive to the environment in aqueous media. 

A recent publication evaluated possible artifacts of FCS measurements and the subsequent effects on the diffusion measurements of enzymes \cite{Gunther2018}. They discussed that enzymes at low concentration can dissociate into smaller subunits. This dissociation cannot be detected by FCS, but would cause an increase of the measured diffusion coefficient. They also described that free dyes remaining in solution can affect the measured autocorrelation functions, which subsequently impact the determination of diffusion rate. Most importantly, they demonstrated that substrate binding to enzymes can cause fluorescence quenching in some cases that resulted in a faster decay of the autocorrelation curves and thus a larger diffusion constant \cite{Gunther2018}. Experts agree that interpretation of autocorrelation curves is complicated and requires modeling of experimental situations. Thus, it is imperative that these results are repeated and recapitulated with a distinct experimental method. Here, we use direct single molecule imaging to visualize the trajectories of enzymes in solution over time and calculate their mean squared displacements (MSD) to determine the diffusion coefficients. This method has the added value that it is truly single molecule and mobility increases were obvious by eye.

\begin{figure}
	\centering
		\includegraphics[width=1\linewidth]{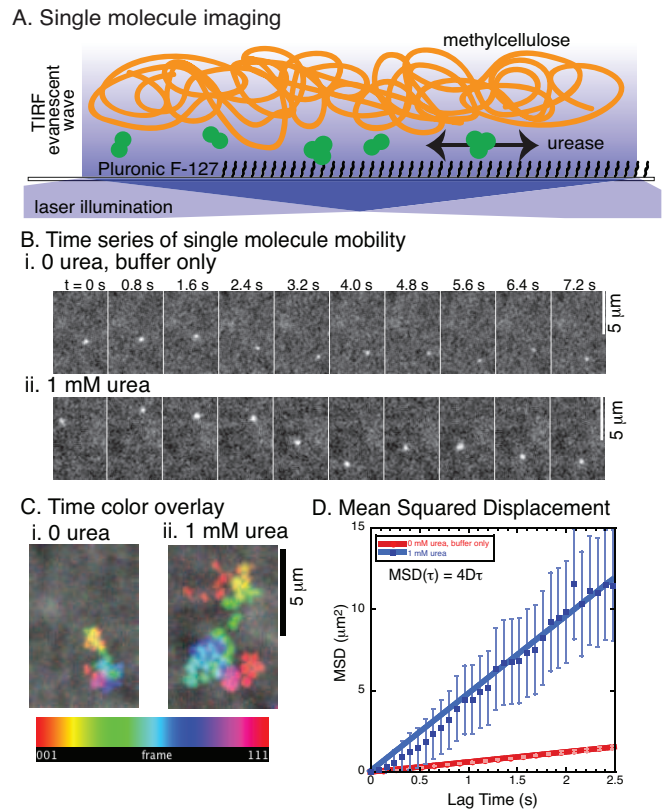}
		\caption{A) Experimental setup for single particle imaging of urease using TIRF imaging of fluorescent urease in a chamber with Pluronic F127 coating the surface and a crowding agent, methylcellulose. B) Example trajectories of a single urease enzyme over time. i) without urea, and ii) with urea at $1~mM$.  Scale bar $5~\mu m$. C) Example 2D trajectories displayed over time as collapsed images with rainbow scale representing time as given in the time color bar over 111 frames with 0.08 s between frames for urease i) without urea, and ii) with $1~mM$ urea. Scale bar $5~\mu m$. D) Calculated mean squared displacement (MSD) plot of the trajectories and fit with a linear equation to determine the diffusion coefficient, $D$, for urease without urea (red circles) and with $1~mM$ urea (blue squares). Error bars represent the standard error of the mean.}
	\label{fig:Method}
\end{figure}

Our single particle tracking experiments were performed with total internal reflection fluorescence (TIRF) microscopy (Fig. \ref{fig:Method}A) using a custom-built laser system (488 nm, 638 nm) constructed around a Nikon Ti-E microscope with a 60x, 1.49 TIRF objective and 2.5x magnification prior to the EM-CCD camera (Andor). We directly observe the diffusing trajectory of each individual enzyme by recording at 8 - 20 frames/s (Fig. \ref{fig:Method}B,C). Enzymes are blocked from sticking to the silanized hydrophobic coverglass by adding Pluroinc F127 block-copolymer (see Supplemental Information). The lifetime of the fluorescence is extended by adding glucose oxidase and catalase as an oxygen scavenging system, which is exactly the same for all experiments. TIRF microscopy can only image the first 300 nm distance from the coverglass (Fig. \ref{fig:Method}A), so all experiments included the addition of methylcellulose to crowd the enzyme close to the surface. Trajectories of enzymes were analyzed by an ImageJ/FIJI plugin ParticleTracker 2D/3D \cite{Sbalzarini2005}(Fig. \ref{fig:Method}D). The diffusion coefficient, $D$ in $m^2/s$, was determined from the slope of the MSD plot according to the Brownian motion equation in 2D: $\big<(\Delta r)^2\big>=4D\tau$, where $\tau$ is the lagtime in $s$. The enzyme we used was urease from Jack Bean (TCI Chemicals), a fast, highly exothermic enzyme, that breaks down its substrate, urea, into ammonia and carbon dioxide. Urease is a hexamer which we fluorescently labeled one fluorophore per monomer with a commercially available dye labeling kit (Thermo Fisher). 

\begin{figure*}
	\centering
		\includegraphics[width=1\linewidth]{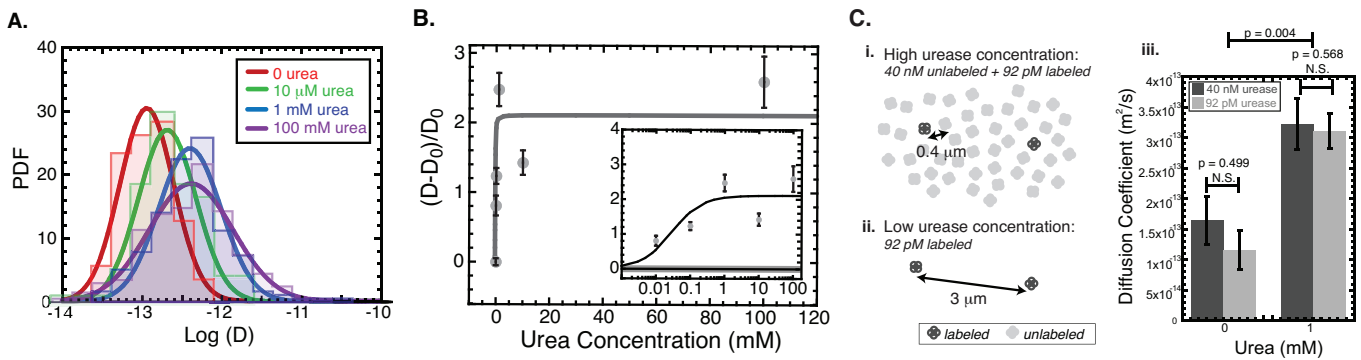}
		\caption{A) Representative probability distribution histograms of log-transformed diffusion data at different urea concentrations $0$ (red region, N = 141), $10\mu M$ (green region, N = 97), $1~mM$ (blue region, N=178), $100~mM$ (purple region, N = 203). Gaussian fit lines $0$ (red line), $10~\mu M$ (green line), $1~mM$ (blue line), $100~mM$ (purple line). Fit parameters can be found in the supplemental information. B)  The normalized relative increase in diffusion coefficient $(D-D_0)/D_0$, plotted as a function of the urea concentration. Inset shows the same data plotted on a logarithmic scale. Error bars are determined from the standard derivations $\sigma$ of the Gaussian distribution fits from part (A). The fit equation is a hyperbolic function with a amplitude and characteristic concentration, $K$; fit parameters given in supplemental information. C) i) Cartoon of $40~nM$ urease with average spacing between molecules of $400~nm$. ii) Cartoon of $90~pM$ urease with average spacing between molecules of $3~\mu m$. iii) Median diffusion coefficients of urease without urea in high urease concentration ($40~nM$, dark gray bars, N = 31) and low urease concentration ($90~pM$, light gray bars, N = 30) and with $1~mM$ urea in high urease concentration ($40~nM$, dark gray bars, N = 35) and low urease concentration ($90~pM$, light gray bars, N = 36). Error bars are determined from the standard derivations $\sigma$ of their Gaussian distribution fits. }
	\label{fig:Substrate}
\end{figure*}

In agreement with prior work, we find that urease displays enhanced diffusion in the presence of its substrate urea. The change in motility was visible directly from trajectories and the MSD plots (Fig. \ref{fig:Method}B-D). For our assays, we measured over 100 single particle trajectories for each experimental condition to obtain statistically significant data. Diffusion data displayed a lognormal distribution that could be plotted and fit with a Gaussian after log-transformation (Fig. \ref{fig:Substrate}A). The mean of the Gaussian fits was then transformed back and used as the general diffusion coefficient for each case (see Supplemental Information for fits).

Interestingly, we find that the relative increase of the diffusion coefficient in our single molecule experiments is significantly higher than those previously reported using FCS methods \cite{Muddana2010,Bustamante}. For the highest concentration of urea we tested ($100~mM$), we found a $\sim 3$ fold increase in the diffusion constant (Fig. \ref{fig:Substrate}B), compared to prior results that showed only a $\sim 30\%$ increase \cite{Muddana2010, Bustamante}. Control experiments performed with green fluorescent protein and inhibited urease that cannot interact with urea both show a slight decrease in diffusion coefficient in the group with the presence of urea (Supplemental Information, Figs. S1-3). These controls demonstrate that the enhanced diffusion of urease is not due to the presence of urea in solution, but rather to the interaction between urea and urease. 



We calculate and plot the relative increase in the diffusion coefficient as a function of urea concentration (Fig. \ref{fig:Substrate}B), which displays a typical hyperbolic dependence of the form: $(D-D_0)/D_0 = A\times \frac{[urea]}{[urea]+K}$, where $D$ is the measured diffusion coefficient, $D_0$ is the diffusion coefficient in the absence of substrate, $A$ is an amplitude, $[urea]$ is the urea substrate concentration, and $K$ is a rate constant. We find that the best fit has $K = 21 \pm 16~ \mu M$ (see Supplemental Information for fit parameters). The $K_D$ for urea binding to urease was reported as $250~\mu M$ \cite{Tanis1968}, and the $K_M$ was reported as $3~mM$ \cite{Krajewska2009}. Comparing our results to these two rate constants, we find that our data is more similar to the binding coefficient,$K_D$, instead of the reaction turn-over rate, $K_M$. Several theoretical models have suggested that substrate binding could change the size or flexibility of enzymes, driving the difference in the diffusion coefficient \cite{Golestanian2017, Illien2017}, but no model has predicted such a large shift in the diffusion coefficient as we measure here.

Prior works have noticed a correlation between the diffusion coefficient increase and the heat released during enzymatic turnover \cite{Bustamante}. Assuming the enzyme size does not change during the turnover, in order for the diffusion coefficient to increase by a factor of 3, as we observe (Fig. \ref{fig:Substrate}B), the temperature would need to increase by $55 ~K$ locally. This increase was estimated by using the Stokes-Einstein relation: $D = \frac{k_BT}{6 \pi \eta a}$, in which the viscosity, $\eta$ in $Pa \cdot s$, is also considered as a function of temperature: $\eta(T) = 2.4 \times 10^{-5} \times 10^{247.8/(T-140)}$ for water \cite{Al-Shemmeri2012} in our calculation (see Supplemental Information for details). Using a rough estimation method as described previously \cite{Bustamante}, the maximum temperature increase within a $1~nm$ water shell around the enzyme would be $0.09~K$ for urease, which is too small to account for the factor of 3 increase in diffusion we observed.

We also estimate the temperature increase around a single enzyme using the solution to the heat diffusion equation with a instantaneous point source. Since the concentration of enzyme was set to be extremely low ($ \sim 100 pM$), each single enzyme is modeled as an instantaneous point source of heat during each enzymatic turnover. Thus, we have:
\begin{equation}
   \Delta T (r,t) = \frac{\Delta Q}{\rho c (4 \pi \kappa t)^{3/2}} \exp {\big[-\frac{r^2}{4 \kappa t}\big]} , \label{heat}
    \label{eqn:heateqn}
\end{equation} 
where $\Delta Q = 25k_BT$ is the heat released from a single catalytic reaction. The background material is water with specific heat capacity $c = 4.18~J/{(K \cdot g)}$, density $\rho = 1~g/cm^{3}$, and thermal diffusivity $\kappa \simeq 10^{-7}~m^{2}/s$. We estimated the temperature increase during one catalytic turnover, with $t = t_{c} = 1/k_{cat} \simeq 10^{-4}~s$ for urease at saturating urea concentrations and used a distance comparable to enzyme size with $r = 2~nm$. We found the temperature shift would be minuscule, $\Delta T \sim 10^{-11}~K$, so it seems unlikely that heating the local environment alone could cause such a large increase in the diffusion coefficient.

Another model suggested that enhanced diffusion could arise from heating the entire chamber due to many enzymes in solution \cite{Golestanian2015}. Using their model, with the parameters of our experiments, we estimated that the temperature increase in the whole chamber would be $\Delta T_{total} \sim 10^{-6}~K$ (see Supplemental Information for information on this estimate), which is still too small to account for the large increase in diffusion coefficients. 
In the above estimations, the enzymes each act as independent sources of heat or activity. Two recent models have taken collective effects of many enzymes into account. One is a collective heating model \cite{Golestanian2015} and another is a collective hydrodynamics model \cite{Mikhailov2015}. Both of these models predict that the diffusion rate increase will depend linearly on the total concentration of the enzymes in solution. 

To test the predictions of these collective models, we repeat our experiments at two different total enzyme concentrations, $40~nM$ and $90~ pM$ (Fig. \ref{fig:Substrate}C). For both groups, we keep the concentration of labeled enzyme constant at single molecule level ($90~pM$). The average spacing between enzymes depends on their concentrations in solution, which we estimate to be $\sim 400~nm$ for $40~nM$ and $\sim 3 ~\mu m$ for $90~pM$  (Fig. \ref{fig:Substrate}Ci-ii). We compared the diffusion coefficients for different concentration groups in the absence of urea or with $1~mM$ urea (saturating concentration Fig. \ref{fig:Substrate}B). We find no difference in the diffusion constants between $40~nM$ and $90~pM$ concentrations for either the buffer case or urea case (Fig. \ref{fig:Substrate}Ciii). Although the proportional relationship with total enzyme concentration was not observed in our experiments, it is possible that collective phenomena would come into play at much higher, non-physiological concentrations of enzymes. Regardless, these collective models cannot explain the 3-fold increase in diffusion that we observe in our experiments. 


As described above, diffusion coefficients can also be significantly altered due to the dissociation of enzyme complexes at the low concentrations used in FCS studies \cite{Gunther2018}. 
 Suppose an enzyme with radius $R$ undergoes a change in size, $\delta R$, during its interaction with the substrate, and the liquid viscosity remains the same. From the Stokes-Einstein equation, the relative change in diffusion can be written as  
\begin{equation}
    \frac{\Delta D}{D_0} = \frac{1}{1+\frac{\delta R}{R}}\frac{T}{T_0}-1 . \label{equ1}
    \label{eqn:equation1}
\end{equation} 
A positive change in $\Delta D$ requires a negative change in $\delta R$, as expected. We can then estimate the required size change of urease in our experiments needed to account for a 3-fold increase in diffusion. For our experiments, $\frac{\Delta D}{D_0} \sim 2$ and $\frac{T}{T_0} \simeq 1$ from the calculations above. We estimate that $\delta R \simeq -\frac{2}{3}R$, corresponding to a 67\% loss of radius. Considering the possibility that enzyme multimers might dissociate at low concentration, the large increase in our diffusion measurements would most likely be due to the dissociation of urease hexamers to smaller oligomers after interacting with urea. 

\begin{figure}
	\centering
		\includegraphics[width=1\linewidth]{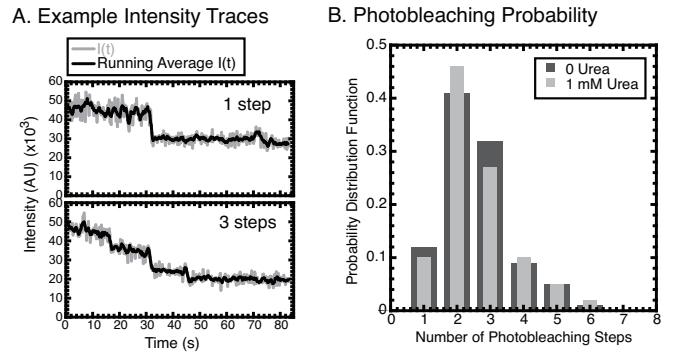}
		\caption{A) Two examples of the intensity of fluorescent urease complexes photobleaching over time, showing a one-step bleach (top) and a three step bleach (bottom). B) The distributions of photobleaching steps directly report the number of fluorescent urease monomers in each complex in the presence of 0 urea (dark gray bars, N = 100) and 1 mM urea (light gray bars, N = 100).}
	\label{fig:bleaching}
\end{figure}

Although, this dissociation process cannot be detected by FCS, it can be directly monitored using our single molecule imaging method. To directly test the oligomerization state of the urease multimers, we perform single molecule photobleaching experiments that reveal the number of urease monomers within each fluorescent complex \cite{Ross2006,Conway2012}. Each urease monomer is covalently labeled with one fluorophore, and there are reported to be 6 monomers per urease complex. We first mix the labeled urease hexamers with urea at $0$ or $1~mM$ concentration allowing them to react and then affix them to the coverglass. Binding to the glass stabilizes their state and makes the local laser illumination and z-height constant for the entire measurement. We use TIRF microscopy to image the enzymes without oxygen scavenging enzymes, so that the fluorophores photobleach over time (Fig. \ref{fig:bleaching}A). 

We count the number of photobleaching steps for each complex, which corresponds to the number of monomers in each complex, and create a histogram of the number of bleaching events (Fig. \ref{fig:bleaching}B). Urease complexes never display more than 6 bleach steps, indicating that the hexamer is the largest oligomerization state. We find that two or three monomers per complex is the most common state for both $0$ and $1~mM$ urea conditions. If the dissociation really occurred due to the presence of urea, we would expect to see a large shift in the distribution of the $1~mM$ urea group to lower numbers of bleaching steps. However, we find no difference between these two distributions according to the Kolmogorov-Smirnov statistical test ($p = 1.0$). From these results, the enhanced diffusion we observed cannot be caused by changes in the oliomerization state. This result also demonstrates another strength of the direct imaging technique we employ over FCS measurements.



In conclusion, we used a distinct method to measure the diffusion of enzymes to test if the enhanced diffusion previously reported was genuine or an artifact of the fluorescence correlation spectroscopy technique. Excitingly, we have verified that the enhanced diffusion of urease occurs on a truly single molecule level. We also observe a higher increase in diffusion rates, by a factor of three, in comparison with the $\sim 30\%$ increase measured with FCS. We find our large increase in diffusion is difficult to account for based on any current physical models based on heat or collective interactions. Finally, the single molecule imaging techniques are able to directly measure the oligomerization state of the enzymes, excluding the possibility that the enhancement in diffusion we observe is caused by the dissociation of enzyme multimers. We expect the direct imaging technqiue will be a powerful, complementary method to test the predictions of future models of the mechanism behind the enhanced diffusion of enzymes.   

\textit{\textbf{Acknowledgements}} MX was partially supported by NSF MRSEC DMR-1420382 to Seth Fraden (Brandeis University) and UMass Faculty Research Grant to JLR. JLR was partially supported by DoD ARO MURI 67455-CH-MUR to S. Thayumanavan. We want to thank Ramin Golestanian for helpful feedback about our manuscript.


\bibliographystyle{apsrev4-1} 
\bibliography{cites.bib} 


\end{document}